\documentclass[a4paper, amsfonts, amssymb, amsmath, reprint, showkeys, nofootinbib, twoside]{revtex4-1}
\usepackage[english]{babel}
\usepackage[utf8]{inputenc}
\usepackage[colorinlistoftodos, color=green!40, prependcaption]{todonotes}
\usepackage{amsthm}
\usepackage{mathtools}
\usepackage{physics}
\usepackage{xcolor}
\usepackage{graphicx}
\usepackage[left=23mm,right=13mm,top=35mm,columnsep=15pt]{geometry} 
\usepackage{adjustbox}
\usepackage{placeins}
\usepackage[T1]{fontenc}
\usepackage{lipsum}
\usepackage{csquotes}
\usepackage{dcolumn}
\usepackage{bm}

\newcommand{\be}{\begin{equation}}       
\newcommand{\ee}{\end{equation}}       
\newcommand{\beqn}{\begin{eqnarray}}
\newcommand{\as}{\alpha_s}
\newcommand{\tb}{\bar{\tau}}
\newcommand{\ab}{\bar{\alpha}}
\newcommand{\hq}{\hat{q}}
\usepackage[pdftex, pdftitle={Article}, pdfauthor={Author}]{hyperref} 
\begin{document}
\title{Hard gluon evolution in the last stage of the bottom-up thermalization}

\author{F. G. Ben}
    \email[Correspondence email address: ]{felipe.gregoletto@ufrgs.br}
\author{M. V. T. Machado}
    \affiliation{Universidade Federal do Rio Grande do Sul, Institute of Physics, Porto Alegre, RS, Brazil}
\date{\today} 

\begin{abstract}
The study of how fast thermalization in heavy ion collisions occurs has been one of the central topics in the heavy ion community. In the weak coupling picture this thermalization occurs from ``the bottom up'': high energy partons, formed early in the collision, radiate low energy gluons which then proceed to equilibrate among themselves, forming a thermal bath that brings the high energy sector to equilibrium. In this scheme we apply a model on parton energy loss to discuss the effects of medium expansion on the thermalization problem and estimate the average transverse momentum diffusivity for thermalization in a Bjorken expanding medium.
\end{abstract}


\maketitle
\section{Outline} \label{sec:outline}

Since the development of high energy particle accelerators, such as RHIC and LHC, particle physics has entered a new era, in which enough energy is available in the collision that a new form of QCD matter, the Quark Gluon Plasma (QGP), is formed during the very first stages of a high energy heavy ion collision (HIC). 
Different stages of such HIC present different degrees of freedom as the QGP expands and evolves, and therefore may be described by different effective theories.

Prior to and right after the collision, the incoming nuclei are formed by highly dense and anisotropic gluonic matter, which leads to collective behavior that's better suited to be described by fields than by particles.
The Colored Glass Condensate (CGC) effective theory has been the standard approach to describe the system at these early stages \cite{gelis2010color,gelis2013color,gelis2014initial}.
As the QGP matter formed after the collision expands and cools down, however, phase-space densities go down and the concept of particles becomes meaningful again. Should these particles be weakly coupled, as the asymptotic freedom of QCD may imply, their evolution is well described by kinetic theory, which is an effective theory that emerges from the assumption that the mean free path between successive interactions is larger than any other microscopic scale \cite{kurkela2015isotropization,kurkelaa2016initial}.

Intermediate stages of a HIC, on the other hand, have been successfully described by relativistic hydrodynamics, both viscous and ideal. Hydrodynamics is a macroscopic theory based on general conservation laws that is able to the describe bulk properties of the system through very few parameters, such as pressure and energy-density \cite{baier2008relativistic,ollitrault2008relativistic}. Multi-particle correlations, with wavelengths much larger than microscopic wavelengths, have been well described using hydrodynamics \cite{heinz2013collective}.
Relativistic hydrodynamics relies on a few postulates that must be fulfilled -- namely, the system should be (very close to) isotropic at the time that hydrodynamics assumes the evolution. In other words, the system should be close to local thermal equilibrium at the time hydrodynamics is initiated. 
In order to reproduce experimental results, however, hydrodynamic simulations have to be started at very early times in the evolution, of order $\tau_{hydro} \sim 1$ fm/c after the collision \cite{kurkelaa2016initial}.
That implies that the conditions for hydrodynamics, in particular local thermal equilibrium, must be reached at very early times in the evolution. 

Understanding how the out of equilibrium QGP formed in the collision, at proper time $\tau = 0$, is able to redistribute energy (competing with its expansion) and quickly thermalize has been one of the central topics in the heavy ion community for the past few years.
Although some evidences point in the direction of a strongly coupled QGP, such as the small value of the $\eta/s$ viscosity to entropy ratio, calculations in this scenario are highly challenging. Promising results in the strong coupling limit have been found through the AdS/CFT correspondence, but are not directly applicable to QCD \cite{chesler2009horizon}.
In the weak coupling scenario, although the CGC effective theory is able to provide a good description of the initial state, it is not able, at leading order (LO), to bring the system to equilibrium. In fact, in a uniformly expanding medium, CGC at LO predicts that the system should remain anisotropic indefinitely \cite{epelbaum2013pressure}. Many different approaches are possible to proceed, such as the study of NLO corrections to the CGC model or the study of anisotropic hydrodynamics \cite{martinez2010dissipative,martinez2011non}, which may be able to relax the requirement of local thermal equilibrium.

Another weak coupling approach that has presented interesting results in the past few years is to use an intermediate effective theory between the initial and the hydro-dynamical state. That intermediate theory has the role of bringing the initial CGC state to a state of local thermal equilibrium, and from that point on hydrodynamics is able to take on the evolution. As mentioned before, as the QGP evolves and its phase-space density goes down, the system may be described by particles through kinetic theory. One approach, then, is to use the CGC theory to provide the initial conditions to kinetic theory, which then is able thermalize the system. This approach has presented promising results and will be used in this manuscript. 

Through kinetic theory, in the weakly coupled particle description, thermalization occurs "from the bottom-up", as presented in the seminal paper in Ref. \cite{baier2001bottom}.
The thermalization can be divided in three different stages. In the first stage, the initial high-energy hard gluons from the collision irradiate soft gluons through bremsstrahlung. These soft gluons carry only a small fraction of the initial gluons momenta and energy, but rapidly increase in number, interact and equilibrate among themselves in the second stage. From that point on, the soft gluons form a thermal bath, and the system is formed by a small number of hard gluons, which carry most of the energy of the system, moving through a soft thermal bath. 

In the third stage of the bottom-up process the interaction of the hard gluons with the soft medium is responsible for depositing the hard sector's energy into the thermal bath, eventually bringing the system to a thermal state. The physics of this stage is the same as that of ``jets'' of high momentum travelling through a thermal medium \cite{schlichting2019first}, which will provide the tools that will be used throughout this paper. The last stage of the bottom-up thermalization also sets the scale for the thermalization time $\tau_{hydro}$, since the previous stages take parametrically less time \cite{kurkela2011bjorken}.

In this work we focus on studying how general features of the hard gluon distribution evolve in the third stage of the bottom-up thermalization. Our study relies on numerical solutions to the evolution equation for the gluon distribution introduced in \cite{blaizot2013medium,blaizot2014probabilistic}, with the inclusion of how the medium expansion modifies the gluon emission rate \cite{arnold2009simple,adhya2020medium}. We are able to estimate how the hard gluon spectra and its energy density evolve over time in the particular case of the thermalization problem, and due to the phenomenological constraint $\tau_{hydro} \sim 1$ fm/c we are able to estimate the rate for transverse momentum broadening due to interactions with the soft thermal bath, analogous to the jet-quenching parameter in jet studies.\\

\section{Evolution of the hard gluon distribution} \label{sec:develop}
In this section we will describe how the distribution of hard gluons evolves as it travels through the soft thermal bath during the final stage of the bottom-up thermalization. We recall that at this time, not very long after the collision, the system is formed by a large number of soft gluons, that have reached equilibrium among themselves, and by a small number of hard gluons. We will assume that at the time the hard gluons were produced the system was in the saturation scenario, of momentum scale $Q_s$, and that, as the third stage of the bottom-up sets in, the hard gluons still carry an energy of order $E\sim Q_s$. They travel through the thermal bath of soft gluons that carry only a small fraction of the system's energy. As noted in the introduction, the interaction of the hard sector with the soft medium will be responsible for depositing most of the system's energy into the thermal bath, bringing the system to a state of local thermal equilibrium in a time scale of order $\tau_{hydro}$. 

We will assume that the soft medium is characterized by a uniform average temperature $T$, therefore neglecting spatial variations on $T$. It is not required, however, for $T$ to be constant in time. In fact, $T$ is related to the density of soft gluons $n_s$ through $n_s \sim T^3$ \cite{baier2001bottom}, such that $T(\tau)$ may be derived from the evolution of $n_s$. For a static medium, as the one described in \cite{blaizot2014probabilistic}, both $n_s$ and $T$ are indeed constant. Although hard gluons near the boundary of the medium, for instance, may deposit less energy to the medium than those in the central region, we will assume that all partons lose energy equally and into the medium.
We will also assume that the high energy partons travelling through the medium do not alter $T$ in a significant way. In Ref. \cite{baier2001bottom}, it was derived that the temperature of the soft thermal bath actually increases linearly with time during thermalization. However, it is estimated that the final temperature is larger than the initial one only by a factor of $\alpha_s^{-1/10}$. We will include the effect of the temperature evolution in a future work \cite{ben2021hard}.

The main mechanism of energy loss for a parton of high momentum travelling through a QCD medium is in-medium bremsstrahlung, subject to the Landau-Pomeranchuk-Migdal (LPM) suppression, which leads to the BDMPS-Z distribution \cite{baier1997radiative,zakharov1997radiative}. For a thermal medium this was analysed in Ref. \cite{jeon2005energy}, that confirmed that the final distribution is governed by the small number of high energy partons travelling through the medium. Subsequent papers \cite{blaizot2013medium,blaizot2014probabilistic} have studied how an initial distribution of hard gluons evolves in a thermal and static QCD medium. In order to perform a similar analysis in the context of the thermalization problem, one has to include the effects of the medium expansion, which will be discussed shortly.

Let us proceed by briefly summarizing the results in Refs. \cite{blaizot2013medium,blaizot2014probabilistic} in order to create the set up for the next sections.  Consider the hard gluon distribution $D(x,\tau) \equiv x \dfrac{dN_g}{dx}$, made of $N_g$ gluons of initial energy $E \sim Q_s$. The energy $\omega$ of a given gluon is represented by the fraction of the initial energy $x = \omega/E$. In the scenario of a hard gluon travelling through a dense thermal medium, the mechanism by which it will lose energy is that of democratic branching: first, it emits a particle of small energy $\omega_{br} \ll E$ that lies in the LPM-suppressed region, such that by a time of order $\tau$ it will split into two gluons of comparable momenta.
Those daughter gluons will cascade further, depositing their energy into the thermal bath of $x \sim T/E$, where $T$ stands for the temperature \cite{baier2001bottom}.
Following the approach on \cite{adhya2020medium}, the proper time $\tau$ will be re-scaled as the dimensionless variable
\be
\tb \equiv \bar{\alpha}\sqrt{\frac{\hat{q_0}}{E}}\tau,
\label{eq:taubar}
\ee
where $\ab = \as N_c /\pi$ and $\hat{q}$ is the transverse momentum broadening rate $\hat{q} \equiv dk_{\perp}^2/dt$, or jet quenching parameter. In anticipation for the case in which the medium expands uniformly, where $\hat{q} = \hat{q}(\tau)$, we will use the notation $\hq_0 = \hq(\tau_0)$, where $\tau_0$ is the time at which our dynamics sets in. For a static medium $\hq = \hq_0$ is a constant.

Considering the probability of a gluon of energy fraction $x$ to split into two gluons that carry a fraction $z$ and $(1 - z)$ of the parent gluon's energy, between times $\tb$ and $\tb + d\tb$, it was derived in Refs. \cite{blaizot2013medium,blaizot2014probabilistic} an evolution equation for the gluon distribution,
\be
\frac{\partial D(x,\tb)}{\partial \tb} = \int \mathcal{K}(z,\tb) \bigg[\sqrt{\frac{z}{x}} D(\frac{x}{z}, \tb) - \frac{z}{\sqrt{x}}D(x,\tb)\bigg] \mathrm d z.
\label{eq:evol}
\ee
The kernel function $\mathcal{K} (z, \tb)$ is related to the emission spectrum $I(z,\tb)$ of the parton through
\be
\bar{\alpha} \mathcal{K}(z,\tb) = \frac{\mathrm{d}I}{\mathrm{d}z \mathrm{d}\tb},
\ee
and will contain the characteristics of the medium and its expansion \cite{arnold2009simple, adhya2020medium}. The factor of $\ab$ was extracted in order to make our notation consistent with the one used in Ref. \cite{arnold2009simple}. 

Although Eq. \eqref{eq:evol} was derived considering the medium to be homogeneous and static, it still holds for an expanding medium \cite{caucal2020jet}, provided that the kernel function $\mathcal{K} (z, \tb)$ is treated accordingly. In fact, Ref. \cite{arnold2009simple} presents a remarkable simple and general form for the emission spectrum, such that the time evolution of the medium may be fully modeled by $\hat{q}(\tau)$ and boundary conditions, from which $\mathcal{K} (z, \tb)$ may be obtained for a variety of different scenarios. In the remaining of this section we will analyze the evolution in two particular cases (namely, static and Bjorken expansion). In Section \ref{sec:results} we further discuss its application to the thermalization scheme.

The following initial conditions will be assumed in what follows: at the beginning of the third bottom-up stage $D(x,\tb)$ is accumulated at the initial energy $E \sim Q_s$, with $x = \omega/E = 1$, that is, $D(x, \tb = \tb_0) = \delta(1-x)$. The distribution is normalized to $\int D(x,\tb_0)\, \mathrm{d} x = 1$, and $\tau_0$ represents the beginning of the third bottom-up stage, with $\tb_0$ given by Eq. \eqref{eq:taubar} with $\tau = \tau_0$.

\subsection{Static Medium}

Eq. \eqref{eq:evol} was analysed for a static thermal medium in Ref. \cite{blaizot2013medium}, where it was considered
\begin{eqnarray}
\mathcal{K}(z) &=& \frac{f(z)}{[z(1-z)]^{3/2}} = \mathcal{K}(1-z), \\
f(z) &=& [1 - z(1-z)]^{5/2},
\end{eqnarray}
valid in the limit $z \ll \tb^2$. For $\mathcal{K}(z,\tau)= \mathcal{K}(z)$ independent of time, the distribution $D(x,\tb)$ will depend on time through the difference $(\tb - \tb_0)$, such that $\tb_0$ can be taken to be $\tb_0=0$ for simplicity.

With the additional simplification $\mathcal{K}(z) = \mathcal{K}_0(z) = 1/[z(1-z)]^{3/2}$, that is, by taking $f(z) = 1$, it is possible to solve Eq. \eqref{eq:evol} analytically through Laplace transform techniques  \cite{blaizot2013medium}, which leads to
\be 
D_0(x,\tb) = \frac{\tb}{\sqrt{x}(1 - x)^{3/2}} \mathrm{e}^{\frac{-\pi\tb^2}{1-x}}.
\label{eq:Danal}
\ee
It is interesting that, in the limit of small $x$, the spectrum has the scaling property $D_0 \propto 1/\sqrt{x}$. As pointed out in Ref. \cite{adhya2020medium} and in the next section, that seems to be a general feature of the solution, that remains valid in the case of an uniformly expanding medium.

Another general feature of the dynamics implied from Eq. \eqref{eq:evol} is that the fraction $\mathcal{E}$ of the initial energy contained in the hard spectrum decreases with time. For the analytic solution in \eqref{eq:Danal} one finds
\be
\mathcal{E}_0(\tb) \equiv \int_0^1 D_0(x,\tb) \, \mathrm{d}x = \mathrm{e}^{-\pi\tb^2},
\label{eq:Eanal}
\ee
which implies an exponential decrease in energy. Formally, this apparent violation on energy conservation is related to the singularity at $x = 0$ -- the point $x=0$ may be seen as a ``sink'' where the energy of the higher modes gets deposited. As it was pointed out in Ref. \cite{blaizot2013medium}, that is related to the property of scaling for small $x$: the fact that the spectrum retains its shape with time for small values of $x$ indicates that the energy is not being deposited at any particular point of $x>0$. Physically, energy conservation is not violated: energy is deposited in the thermal bath of $x \sim T/Q_s$. At $\tb = 1$, the energy in the spectrum is less than $0.05$ times its initial value and the properties of the system may be fully characterized by the thermal bath. The system may be treated as thermal once the energy contained in the hard sector is fully deposited in the soft bath, that is, $\mathcal{E}(\tau = \tau_{hydro}) \approx 0$.

\begin{figure*}[!tb]
\centering
\includegraphics[scale=0.7]{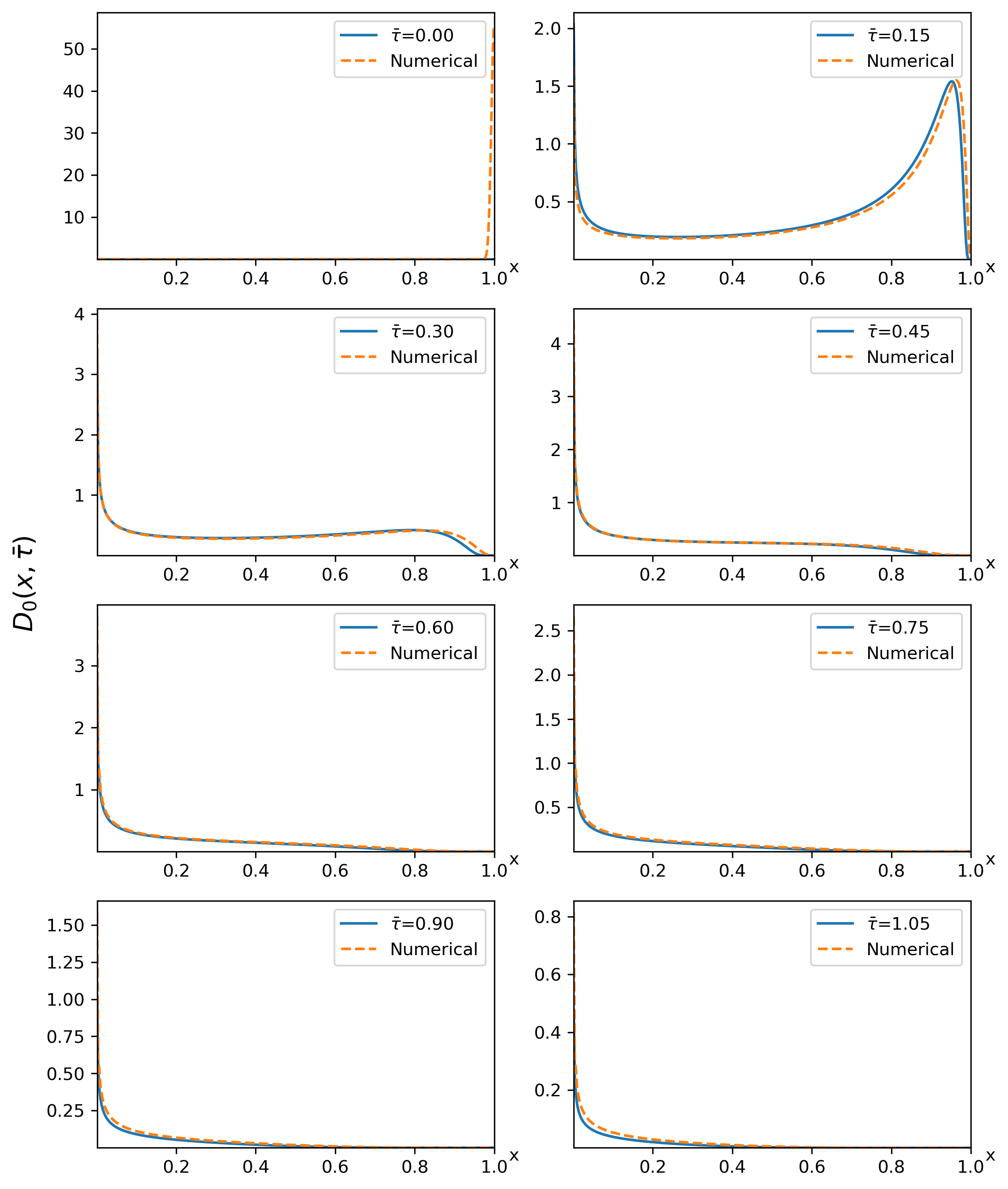}
\caption{Graph of $D_0(x,\tb)$, where $D_0$ is given by Eq. \eqref{eq:Danal}, as a function of $x$ for several values of $\tb$ (continuous line), together with the numerical solution (dotted line).}
\label{fig:Danal}
\end{figure*}

The evolution of $D_0(x,\tb)$ with time is presented in Fig. \ref{fig:Danal}. The spectrum of (initially) hard gluons is accumulated at $x = 1$ in the beginning, and the modes of $0 < x < 1$ are populated with time. The distribution eventually dies off. In order to validate our numerical routine, Eq. \eqref{eq:evol} was also solved numerically for $\mathcal{K}_0$ and is represented alongside the analytic solution (dotted line on the graph). 

\subsection{Uniformly Expanding Medium}

Let us now consider a uniformly expanding medium, which, at first glance, it is expected to be a better description of the situation after the collision (see the following section for an opposing argument). More specifically, we will adopt the model of a uniform longitudinal expansion, initially proposed by Bjorken \cite{bjorken1983highly}. In this model, the medium is considered to be uniformly expanding in the z-direction, with a uniform velocity distribution. Transverse momentum diffusivity $\hq$ is now a function of time (see, for instance, Ref. \cite{iancu2018jet} for a comprehensive discussion):
\be
\hat{q}(\tau) = \hat{q}(\tau_0)  \bigg( \frac{\tau_0}{\tau} \bigg)^{\beta},
\label{eq:qtau}
\ee
with $\beta$ close to $1$ (the limit $\beta \rightarrow 1$ corresponds to an ideal gas).

Following the approach in Ref. \cite{arnold2009simple, adhya2020medium}, in this scenario we have
\be 
\begin{split}
    \bar{\alpha} &\mathcal{K}(z,\tb) = \frac{\as}{2\pi} P_{gg}(z) \kappa (z) \sqrt{\frac{\tau_0}{\tau + \tau_0}} \\
    &\times \mathrm{Re} \bigg[(1-i) \frac{J_{\nu}(z_0)Y_{\nu - 1}(z_L) - Y_{\nu}(z_0)J_{\nu -1}(z_L)}{J_{\nu}(z_L)Y_{\nu - 1}(z_L) - Y_{\nu}(z_L)J_{\nu -1}(z_L)} \bigg],
\end{split}
\label{eq:Kbj}
\ee
where $P_{gg}(z) = 2 N_c [1 - z(1-z)]^2/z(1-z)$ is the Altarelli-Parisi splitting function, $$\kappa (z) \equiv \sqrt{\frac{1-z(1-z)}{z(1-z)}},$$ $J_{\nu}$ and $Y_{\nu}$ represent modified Bessel functions of the first and second kind, respectively, and
\begin{eqnarray}
\nu &\equiv & \frac{1}{2 - \beta}, \\
z_0 &=& \nu(1-i)\kappa(z)\tb_0 ,\\
z_L &=& \nu(1-i)\kappa(z)\tb_0 \bigg( \frac{\tb + \tb_0}{\tb_0} \bigg) ^{1/2\nu},
\end{eqnarray}
where $L \sim \tau$ was considered as the scale for the size of the medium. In the following we will consider the limit $\beta = 1$, which implies $\nu = 1$.

It is worth pointing out that now $\mathcal{K}$ is also a function of the time $\tb_0$, where the dynamics of the third bottom-up stage is initiated\footnote{Parametric estimates of time scales, for different bottom-up stages, were made in the original paper \cite{baier2001bottom} and were revisited, including the role of plasma instabilities, in Ref. \cite{kurkela2011bjorken}.}.
In terms of the physical time $\tau$, $\tau_0$ is some time in the interval $0 \lesssim \tau_0 < \tau_{hydro}$.
Phenomenologically, $\tau_0$ may be treated as a free parameter to be varied in simulations, subject to the restriction $\tau_{hydro} \sim 1$ fm/c.
The thermalization time scale, in this framework, may be analysed by the fraction of the initial energy contained in the distribution $D(x,\tb)$,
\be
\mathcal{E}(\tb) \equiv \int_0^1 D(x,\tb) \, \mathrm{d}x = \int_0^1 x \frac{dN_g}{dx}  \, \mathrm{d}x ,
\label{eq:energia}
\ee
since $\mathcal{E}(\tau = \tau_{hydro}) \approx 0$, as discussed previously.

Eq. \eqref{eq:evol} was solved numerically for $\mathcal{K}$ given by \eqref{eq:Kbj}. The solution is valid for $\tb > \tb_0$, and $\tb_0 = 0.02$ was adopted in our model. It is worth pointing out that the value of $\tb_0$ depends not only on the value of $\tau_0$ but also on the value of $\hat{q}_0$. Those values may be motivated by parametric estimates or may be treated as free parameters that can be adjusted to fit the time scale $\tau_{hydro}$. In order to reduce the number of degrees of freedom in our analysis, the value of $\tb_0$ will be taken as fixed. Although it is expected that the general features of the distribution do not depend on the unphysical choice of $\tb_0$, its influence in the results may be presented, for instance, by varying its value and treating it as a systematic error in simulations.

\begin{figure}[!tb]
\centering
\includegraphics[scale=0.6]{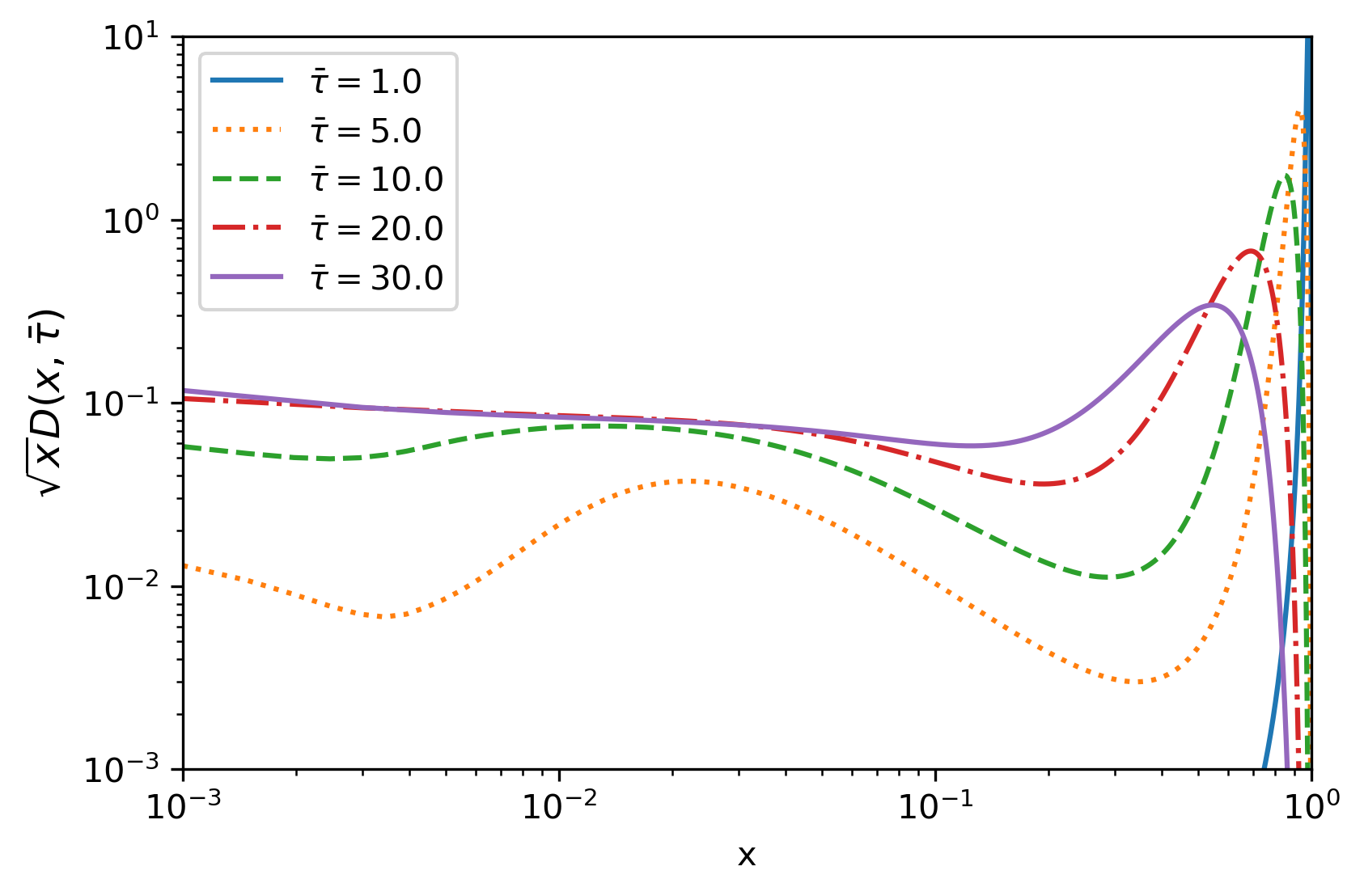}
\caption{Graph of $\sqrt{x}D(x,\tb)$ for a longitudinally uniformly expanding medium, for different values of $\tb$, with initial conditions $D(x,\tb_0) = \delta (1-x)$ and $\tb_0 = 0.02$.}
\label{fig:RaizXDBjorken}
\end{figure}

\begin{figure*}[!tb]
\centering
\includegraphics[scale=0.7]{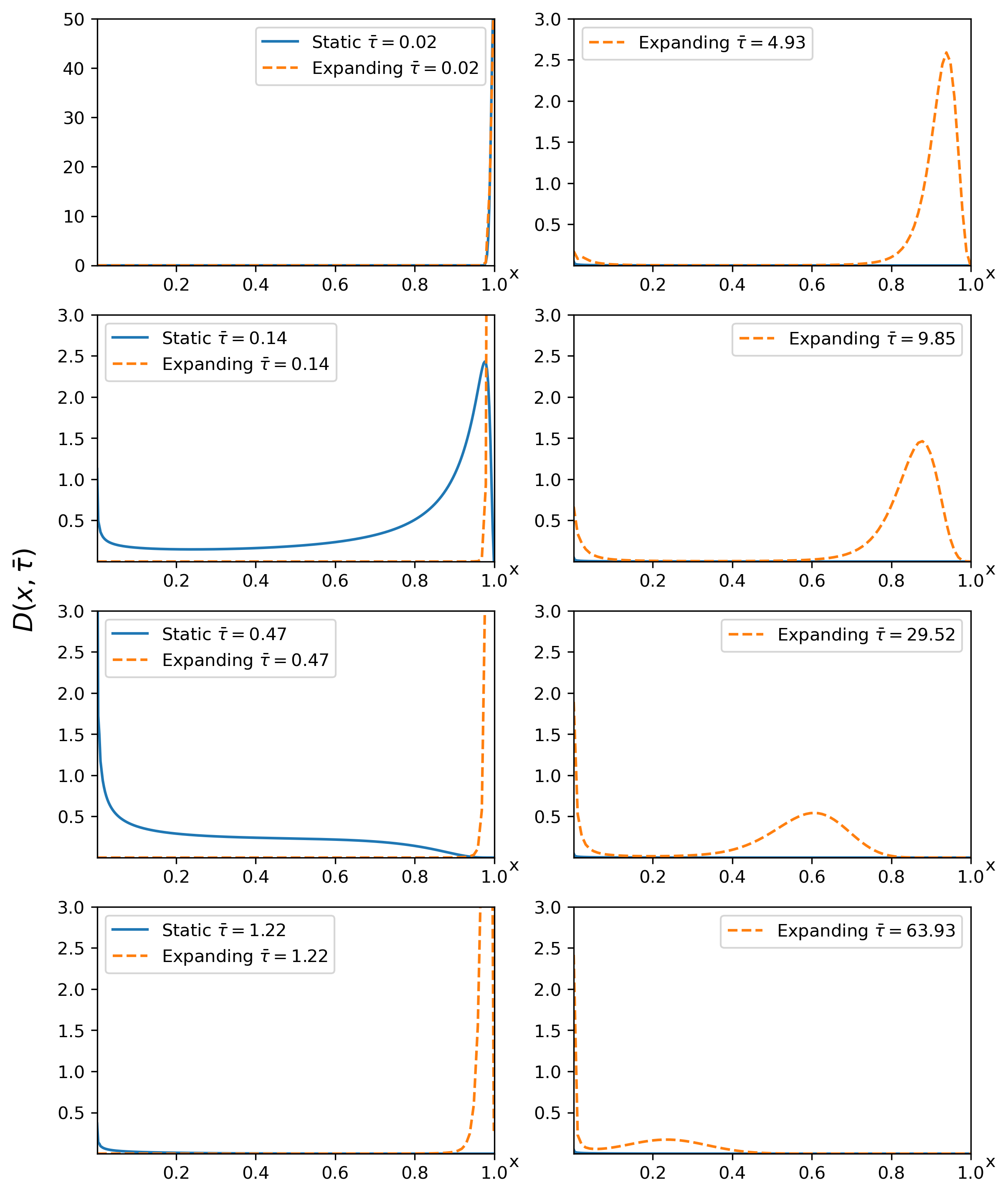}
\caption{Graph of $D(x,\tb)$ for a uniformly expanding medium in the z-direction (dotted line) for several values of $\tb$. For comparison, it is also shown the distribution for the static case, $D_0(x,\tb)$ (continuous line).}
\label{fig:ComparacaoDistribuicoes}
\end{figure*}

An important feature of $D(x,\tb)$ in the case of a uniformly expanding medium is that the scaling $D(x,\tb) \propto 1/\sqrt{x}$, that was motivated by the analytical solution with a simplified kernel $\mathcal{K}_0$, still holds for sufficiently small values of $x$, in spite of the medium expansion. That is illustrated in the behavior presented in Fig. \ref{fig:RaizXDBjorken}.
As pointed out previously and in Ref. \cite{blaizot2014probabilistic}, the scaling $D_0(x,\tb)/\sqrt{x} \sim$ constant for small values of $x$, that is verified in the analytic solution to the static case, and the fact that the distribution retains its shape for small $x$ as time evolves, suggests that energy is not being deposited in any particular value of $x$, that is, it is flowing to the thermal bath. We see, in fact, that for the static case the distribution is soon ``washed out'' in Fig. \ref{fig:Danal}, although we started with a distribution concentrated at $x \sim 1$. In the expanding case, seen in Fig. \ref{fig:RaizXDBjorken}, evolution is slower: although the average value of $x$ decreases with time, the distribution remains populated at larger values of $x$ for a longer time, which is washed out at larger time scales. Not only the flow of energy to the thermal bath is slowed down, but the flow of energy to modes of smaller $x$ is also slower, leading to the ``bumps'' seen in Fig. \ref{fig:RaizXDBjorken} and Fig. \ref{fig:ComparacaoDistribuicoes} at early times.

Fig. \ref{fig:ComparacaoDistribuicoes} presents a comparison between the distribution $D(x,\tb)$ for an uniformly expanding medium (dotted line) and the distribution of gluons in the static case $D_0(x,\tb)$ (continuous line), for different values of $\tb$. Both evolutions share some similar features: the behavior alludes to a source, initially located at $x = 1$, that is dampened as the spectrum propagates in the direction of $x \rightarrow 0$. In the expanding medium, however, evolution is between one and two orders of magnitude slower. In the static case the spectrum fades in $\tb \sim 1$, while it takes $\tb \sim 100$ in the expanding medium.
Qualitatively this behaviour is expected: energy and momentum exchange among partons is significantly slower when competing with the medium expansion, delaying thermalization.

\section{Transverse Momentum Broadening Rate} \label{sec:results}

\begin{figure*}[htp]
\centering
\includegraphics[scale=0.75]{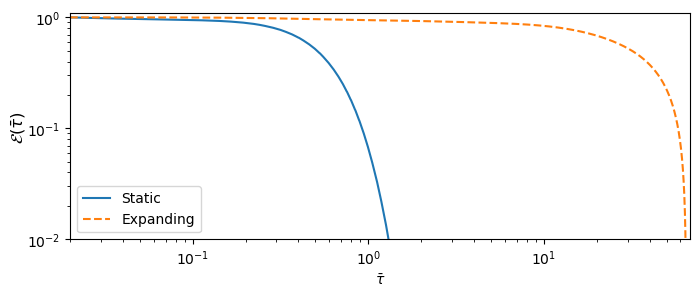}
\caption{Fraction of the initial energy contained in the hard spectrum as a function of $\tb$ for both static (continuous line) and uniformly expanding (dotted line) mediums.}
\label{fig:energiaespectro}
\end{figure*}

Analysis of the energy spectrum provides insights on the thermal bath: energy loss of the hard gluons is dictated by the medium properties. The fraction of the initial energy contained in the spectrum is given by \eqref{eq:energia}. The results for the scenarios discussed in the previous section (static and longitudinal uniform expansion) were evaluated and are presented in Fig. \ref{fig:energiaespectro}. For the static medium, energy decays exponentially in $\tb^2$, according to Eq. \eqref{eq:Eanal}, and falls below 10\% for $\tb \sim 1$.
For the medium in longitudinal uniform expansion it presents almost a linear decay for $\tb \gg \tb_0$, falling below 10\% its initial value for $\tb \sim 65$.

In light of the thermalization problem, the phenomenological constraint $\tau_{hydro} \approx 1$ fm/c $\approx 5$ GeV$^{-1}$, together with Eq. \eqref{eq:taubar}, allows for an estimate on transverse momentum diffusivity $\hq_0$. By use of Eq. \eqref{eq:taubar} we can write
\be
\hat{q_0} = Q_s \bigg(\frac{\tb_{hydro}}{\bar{\alpha}\tau_{hydro}}\bigg)^2.
\ee
Let us take $\tb_{hydro}=1$ for the static case and $\tb_{hydro} = 65$ for the expanding case. Although somewhat arbitrary, it is clear from Fig. \ref{fig:energiaespectro} that those values are enough for an order-of-magnitude comparison between those two scenarios. Considering $Q_s = 2$ GeV and $\as = 0.3$, we find $\hat{q}_{0ST} = 0.95$ GeV$^3$ for the static medium and $\hat{q}_{0EXP} = 4021.33$ GeV$^3$ for the expanding medium.

Considering that in the expanding case $\hat{q}= \hat{q}(\tau)$, according to \eqref{eq:qtau}, $\hq_{EXP}$ reaches $\sim 1.6$ GeV$^3$ in $\tau = \tau_{hydro}$, and we may consider the average value $\langle \hat{q}_{EXP} \rangle$ as defined in \cite{adhya2020medium}:
\be 
\begin{split}
    \langle \hat{q}_{EXP} \rangle &= \frac{2}{(\tau_{hydro}-\tau_0)^2}\int_{\tau_0}^{\tau_{hydro}} (\tau-\tau_0) \hat{q}_{0EXP}  \bigg( \frac{\tau_0}{\tau} \bigg) \, \mathrm{d}\tau \\
    &\approx \frac{2\hat{q}_0\tau_0}{\tau_{hydro}} \approx 2.47 \, \mathrm{GeV}^3,
\end{split}
\ee
that is, $\langle \hq \rangle$ must be, on average, $\sim 2.6$ times larger in a longitudinally uniformly expanding medium than in a static medium in order to guarantee a thermalization time scale compatible with current phenomenology.
It is instructive to compare with $\hq$ in jet quenching studies, in which energetic partons travel through QCD matter in their way to the detectors. Hadronic jets, however, spend most of their trajectory in a medium formed by hydrodynamic matter, which implies $\langle \hat{q}_{JET} \rangle$ is dominated by the dynamics in $\tau \gtrsim \tau_{hydro}$. Typical values are of order $\langle \hat{q}_{JET} \rangle \approx 0.2$ GeV$^3$ \cite{adhya2020medium}, smaller than $\langle \hat{q}_{EXP} \rangle$, as expected for a less dense medium.

\section{Conclusions} \label{sec:conclusions}

In summary, in this work we applied current tools for parton energy loss as it travels through a thermal QCD medium, in the weak coupling scenario, to describe a toy model for the hard gluon evolution in the thermalization problem in a Bjorken-expanding medium. In general, medium expansion contributes to slowing down the thermalization process, as expected, and soft medium properties like $\hq$ may be estimated from this model. In particular, it is shown that $\langle \hq \rangle$ must be larger in a uniformly expanding medium than the value it would have to assume in a static one, in order to ensure thermalization in a time scale compatible with current estimates on $\tau_{hydro}$.
\section*{Acknowledgements} \label{sec:acknowledgements}
This work was financed by the Brazilian funding agency CNPq and in part by CAPES - Finance Code 001.

\bibliography{fgbenbib}


\end{document}